# Efficient Packet Forwarding Approach in Vehicular Ad Hoc Networks Using EBGR Algorithm

**K.Prasanth[1], Dr. K. Duraiswamy[2], K. Jayasudha[3] and Dr. C. Chandrasekar[4]**

**[1]Research Scholar, Department of Information Technology**
**K.S.Rangasamy College of Technology, Tiruchengode – 637 215, Tamilnadu, India**

**[2]Dean Academic, Department of Computer Science**
**K.S.Rangasamy College of Technology, Tiruchengode – 637 215, Tamilnadu, India**

**[3]Research Scholar, Department of Computer Applications**
**K.S.R College of Engineering, Tiruchengode – 637 215, Tamilnadu, India**

**[4]Reader, Periyar University, Salem, Tamilnadu, India**

**Abstract**

VANETs (Vehicular Ad hoc Networks) are highly mobile wireless ad hoc networks and will play an important role in public safety communications and commercial applications. Routing of data in VANETs is a challenging task due to rapidly changing topology and high speed mobility of vehicles. Conventional routing protocols in MANETs (Mobile Ad hoc Networks) are unable to fully address the unique characteristics in vehicular networks. In this paper, we propose EBGR (Edge Node Based Greedy Routing), a reliable greedy position based routing approach to forward packets to the node present in the edge of the transmission range of source/forwarding node as most suitable next hop, with consideration of nodes moving in the direction of the destination. We propose Revival Mobility model (RMM) to evaluate the performance of our routing technique. This paper presents a detailed description of our approach and simulation results show that packet delivery ratio is improved considerably compared to other routing techniques of VANET.

***Keywords:*** *Vehicular Ad hoc Networks, Greedy Position Based Routing, EBGR, Revival Mobility Model, Packet Delivery Ratio.*

## 1. Introduction

Inter-vehicle communication (IVC) is attracting considerable attention from the research community and the automotive industry. It is beneficial in providing intelligent transportation system (ITS) as well as drivers and passenger's assistant services. VANETs is a form of mobile ad hoc network providing communications among nearby vehicles as well as between vehicles and nearby fixed equipment, usually described as roadside equipment. VANETs have similar or different radio interface technologies, employing short-range to medium-range communication systems. The radio range of VANETs is several hundred meters, typically between 250 and 300 meters. In US, the Federal Communications Commission (FCC) has allocated 75 MHz in 5.9 GHz band for licensed Dedicated Short Range Communication (DSRC) for vehicle-to-vehicle and vehicle to infrastructure communications. In Europe, the Car-to- Car Communication Consortium (C2C-CC) has been initiated by car manufacturers and automotive OEMs (original equipment manufacturers), with the main objective of increasing road traffic safety and efficiency by means of inter-vehicle communication.

## 2. Related Work

In this section, we briefly summarize the characteristics of VANETs related to routing and also we will survey the existing routing schemes in both MANETs and VANETs in vehicular environments.

### 2.1. VANETs Characteristics

In the following, we summarize the unique characteristics of VANETs compared with MANETs.

***Unlimited transmission power*:** Mobile device power issues are not a significant constraint in vehicular Networks. Since the vehicle itself can provide continuous power to computing and communication devices.
***High computational capability:*** Operating vehicles can afford significant computing, communication and sensing capabilities.

***Highly dynamic topology:*** Vehicular network scenarios are very different from classic ad hoc networks. In VANETs, vehicles can move fast. It can join and leave the network much more frequently than MANETs. Since the radio range is small compared with the high speed of vehicles (typically, the radio range is only 250 meters while the speed for vehicles in freeway will be 30m/s). This indicates the topology in VANETs changes much more frequently.

***Predicable Mobility:*** Unlike classic mobile ad hoc networks, where it is hard to predict the nodes' mobility, vehicles tend to have very predictable movements that are (usually) limited to roadways. The movement of nodes in VANETs is constrained by the layout of roads. Roadway information is often available from positioning systems and map based technologies such as GPS. Each pair of nodes can communicate directly when they are within the radio range.

***Potentially large scale:*** Unlike most ad hoc networks studied in the literature that usually assume a limited network size, vehicular networks can is extended over the entire road network and include many participants.

***Partitioned network:*** Vehicular networks will be frequently partitioned. The dynamic nature of traffic may result in large inter-vehicle gaps in sparsely populated scenarios and hence in several isolated clusters of nodes.

***Network connectivity:*** The degree to which the network is connected is highly dependent on two factors: the range of wireless links and the fraction of participant vehicles, where only a fraction of vehicles on the road could be equipped with wireless interfaces.

## 2.2 Routing protocols in MANET

The routing protocols in MANETs can be classified by their properties. On one hand, they can be classified into two categories, proactive and reactive.

***Proactive routing algorithms*** employ classical routing strategies such as distance-vector routing (e.g., DSDV [1]) or link-state routing (e.g., OLSR [2] and TBRPF [3]). They maintain routing information about the available paths in the network even if these paths are not currently used. The main drawback of these approaches is that the maintenance of unused paths may occupy a significant part of the available bandwidth if the topology of the network changes frequently [4]. Since a network between cars is extremely dynamic we did not further investigate proactive approaches.

***Reactive routing protocols*** such as DSR [5], TORA [6], and AODV [7] maintain only the routes that are currently in use, thereby reducing the burden on the network when only a small subset of all available routes is in use at any time. It can be expected that communication between cars will only use a very limited number of routes, therefore reactive routing seems to fit this application scenario. As a representative of the reactive approaches we have chosen DSR, since it has been shown to be superior to many other existing reactive ad-hoc routing protocols in [8].

***Position-based routing algorithms*** require information about the physical position of the participating nodes. This position is made available to the direct neighbors in the form periodically transmitted beacons. A sender can request the position of a receiver by means of a location service. The routing decision at each node is then based on the destination's position contained in the packet and the position of the forwarding node's neighbors. Position-based routing does not require the establishment or maintenance of routes. Examples for position-based routing algorithms are face-2 [9], GPSR [10], DREAM [11] and terminodes routing [12]. As a representative of the position based algorithms we have selected GPSR, (which is algorithmically identical to face-2), since it seems to be scalable and well suited for very dynamic networks.

## 2.3. Routing protocols in VANET

Following are a summary of representative VANETs routing algorithms.

### 2.3.1 GSR (Geographic Source Routing)

Lochert et al. in [13] proposed GSR, a position-based routing with topological information. This approach employs greedy forwarding along a pre-selected shortest path. The simulation results show that GSR outperforms topology based approaches (AODV and DSR) with respect to packet delivery ratio and latency by using realistic vehicular traffic. But this approach neglects the case that there are not enough nodes for forwarding packets when the traffic density is low. Low traffic density will make it difficult to find an end-to-end connection along the pre-selected path.

### 2.3.2 GPCR (Greedy Perimeter Coordinator Routing)

To deal with the challenges of city scenarios, Lochert et al. designed GPCR in [14]. This protocol employs a restricted greedy forwarding procedure along a preselected path. When choosing the next hop, a coordinator (the node on a junction) is preferred to a non coordinator node, even if it is not the geographical closest node to destination. Similar to GSR, GPCR neglects the case of low traffic density as well.

### 2.3.3 A-STAR (Anchor-based Street and Traffic Aware Routing)

To guarantee an end-to-end connection even in a vehicular network with low traffic density, Seet et al. proposed A-STAR [15]. A-STAR uses information on city bus routes to identify an anchor path with high connectivity for packet delivery. By using an anchor path, A-STAR guarantees to find an end-to-end connection even in the case of low traffic density. This position-based scheme also employs a route recovery strategy when the packets are routed to a local optimum by computing a new anchor path from local maximum to which the packet is routed. The simulation results show A-STAR achieves obvious network performance improvement compared with GSR and GPSR. But the routing path may not be optimal because it is along the anchor path. It results in large delay.

### 2.3.4 MDDV (Mobility-Centric Data Dissemination Algorithm for Vehicular Networks)

To achieve reliable and efficient routing, Wu et al. proposed MDDV [16] that combines opportunistic forwarding, geographical forwarding, and trajectory-based forwarding. MDDV takes into account the traffic density. A forwarding trajectory is specified extending from the source to the destination (trajectory-based forwarding), along which a message will be moved geographically closer to the destination (geographical forwarding). The selection of forwarding trajectory uses the geographical knowledge and traffic density. MDDV assumes the traffic density is static. Messages are forwarded along the forwarding trajectory through intermediate nodes which store and forward messages opportunistically. This approach is focusing on reliable routing. The trajectory-based forwarding will lead to large delay if the traffic density varies by time.

### 2.3.5 VADD (Vehicle-Assisted Data Delivery)

To guarantee an end-to-end connection in a sparse network with tolerable delay, Zhao and Cao proposed VADD [17] based on the idea of carry and forward by using predicable mobility specific to the sparse networks. Instead of routing along a pre-select path, VADD chooses next hop based on the highest pre-defined direction priority by selecting the closest one to the destination. The simulation results show VADD outperforms GPSR in terms of packet delivery ratio, data packet delay, and traffic overhead. This approach predicts the directions of vehicles movement. But it doesn't predict the environment change in the future.

### 2.3.6 PDGR (Predictive Directional Greedy Routing)

Jiayu Gong proposed PDGR [18], in which the weighted score is calculated for current neighbors and possible future neighbors of packet carrier. With Predictive DGR the weighted scores of immediate nodes 2-hops away are also calculated beforehand. Here next hop selection is done on prediction and it is not reliable at all situations. It doesn't guarantee the delivery of packet to the node present in the edge of the transmission range of forwarding node, which is considered as most suitable next hop, due to high dynamics of vehicles. This will lead to low packet delivery ratio, high end to end delay and increased packet drops.

The various routing protocols of MANETs and VANETs are analyzed and drawbacks of those routing protocols are described in the Table 1.

Table 1
Drawbacks of routing protocols in MANET and VANET

| Routing Protocols | Drawbacks |
|---|---|
| GPSR | Frequent network disconnection.<br>Routing loops.<br>Too many hops.<br>Routing in wrong direction. |
| GSR | End to end connection is difficult in low traffic density. |
| GPCR | End to end connection is difficult in low traffic density. |
| A-STAR | Routing paths are not optimal and results in large delay of packet transmission |
| MDDV | Large delay if the traffic density varies by time. |
| VADD | Large delay due to varying topology and varying traffic density. |
| PDGR | Too many hops.<br>Large delay if the traffic density is high.<br>Low packet delivery ratio.<br>Frequent network disconnection. |

## 3. Proposed Routing Algorithm

### 3.1. Edge Node Based Greedy Routing Algorithm (EBGR)

EBGR is a reliable greedy position base routing algorithm designed for sending messages from any node to any other node (unicast) or from one node to all other nodes (broadcast/multicast) in a vehicular ad hoc network. The general design goals of the EBGR algorithm are to optimize the packet behavior for ad hoc networks with high mobility and to deliver messages with high reliability.

The EBGR algorithm has six basic functional units. First is Neighbor Node Identification (NNI), second is Distance Calculation (DC), third is Direction of Motion Identification (DMI), fourth is Reckoning Link Stability (RLS), fifth is Potential score calculation (PS) and sixth is Edge Node Selection (ENS). The NNI is responsible for collection of information of all neighbor nodes present within the transmission range of source/forwarder node at any time. The DC is responsible for calculating the closeness of next hop using distance information from the GPS. DMI is responsible to identify the direction of motion of neighbor nodes which is moving towards the direction of destination. The RLS is responsible for identifying link stability between the source/forwarder node and its neighbor nodes. The PS is responsible to calculate potential score and identifies the neighbor node having higher potential for further forwarding of a particular packet to destination. The ENS is responsible to select an edge node having higher potential score in different levels of transmission range. In the following section, the general assumptions of EBGR algorithm are briefly discussed and then functional units of EBGR algorithm are discussed in detail.

### 3.2. Assumptions

The algorithm design is based on the following assumptions: All nodes are equipped with GPS receivers, digital maps, optional sensors and On Board Units (OBU). Location information of all vehicles/nodes can be identified with the help of GPS receivers. The only communications paths available are via the ad-hoc network and there is no other communication infrastructure. Node power is not the limiting factor for the design. Communications are message oriented. The Maximum Transmission Range (MTR) of each node in the environment is 250m.

### 3.3. Neighbor Node Identification (NNI)

Neighbor node identification is the process whereby a vehicle/node identifies its current neighbors within its transmission range. For a particular vehicle, any other vehicle that is within its radio transmission range is called a neighbor. All vehicles consist of neighbor set which holds details of its neighbor vehicles. Since all nodes might be moving, the neighbors for a particular mobile node are always changing. The neighbor set is dynamic and needs to be updated frequently. Generally, neighbor node identification is realized by using periodic beacon messages. The beacon message consists of node ID, node location and timestamp. Each node informs other nodes of its existence by sending out beacon message periodically. All nodes within the transmission range of source/packet forwarding node will intimate its presence by sending a beacon message every $\mu$ second. After the reception of a beacon, each node will update its neighbor set table. If a node position is changed, then it will update its position to all neighbors by sending beacon signal. If a known neighbor, times out after $\alpha * \mu$ seconds without having received a beacon ($\alpha$ is the number of beacons that a node is allowed to miss) and it will be removed from the neighbor set table.

### 3.4. Distance calculation (DC)

The location and distance information of all vehicles/nodes can be identified with the help of GPS receivers. It can be communicated to neighbor vehicles using periodic beacon messages. The neighbor node which is closer to the destination node is calculated. The closeness of next hop is identified by the mathematical model [18] and it is shown in Fig.1.

$$\boldsymbol{DC} = \left(\mathbf{1} - \frac{\boldsymbol{D_i}}{\boldsymbol{D_c}}\right)$$

***Here,***
$\boldsymbol{D_i}$ : *Shortest distance from edge node i to destination D.*
$\boldsymbol{D_c}$ : *Shortest distance from packet forwarding node c to destination D.*
$\frac{\boldsymbol{D_i}}{\boldsymbol{D_c}}$ : *Closeness of nexthop.*

Fig. 1 Distance Calculation in EBGR

### 3.5. Direction of Motion Identification (DMI)

The appropriate neighbor node which is moving towards the direction of destination node is identified using the mathematical model [18] and it is shown in Fig.2.

$$\boldsymbol{DMI} = \cos(\vec{\boldsymbol{v}}_{\boldsymbol{i}}, \vec{\boldsymbol{l}}_{\boldsymbol{i,d}})$$

***Here,***
$\vec{\boldsymbol{v}}_{\boldsymbol{i}}$ : *Vector for velocity of edge node i.*
$\vec{\boldsymbol{l}}_{\boldsymbol{i,d}}$ : *Vector for the location of edge node i to the location of destination node D.*
$\cos(\vec{\boldsymbol{v}}_{\boldsymbol{i}}, \vec{\boldsymbol{l}}_{\boldsymbol{i,d}})$ : *Cosine value of angle made by these vectors*

Fig. 2 Direction of Movement Identification in EBGR

The cosine value of vector for velocity of edge node i and vector for location of edge node i to the location of destination node D is measured. A large cosine value implies a vehicle/node can still approach the destination closer and closer along its current direction.

### 3.6 Reckoning Link Stability (RLS)

Each vehicle estimates the Link Stability (LS) for each neighboring vehicle before selecting the next hop for the data forwarding/sending. The LS is a relation between the link communication lifetime and a constant value (say: σ) which represents in general cases the routing route validity time, and it depends on the used routing protocol. Fig.3 shows how link lifetimes are estimated [19] based on neighbors' movement information.

The lifetime of the link (i, j) $LifeTime[i,j]$ corresponds to the estimated time $\Delta t = t_1 - t_0$ with $t_1$ is the time when $D_1$ becomes equal or bigger than the communication range R (i.e. the time when j goes out of the communication range of i). $D_1$ and $\Delta t$ are estimated using the initial positions of i and j $((X_{i0}, Y_{i0})$ and $(X_{j0}, Y_{j0})$, and their initial speeds $\vec{V_i}$ and $\vec{V_j}$ respectively).

$$D_1{}^2 = ((X_{i0} + Vx_i\Delta t) - (X_{j0} + Vx_j\Delta t) + (Y_{i0} + Vy_i\Delta t) - (Y_{j0} + Vy_j\Delta t))^2$$

$$D_1{}^2 = A\Delta t^2 + B\Delta t + C$$

$$A = (Vx_i - Vx_j)^2 + (Vy_i - Vy_j)^2$$

$$B = 2[(X_{i0} - X_{j0})(Vx_i - Vx_j) + (X_{i0} - X_{j0})(Vy_i - Vy_j)]$$

$$C = (X_{i0} - X_{j0})^2 + (Y_{i0} - Y_{j0})^2$$

*Solving the equation*:

$$A\Delta t^2 + B\Delta t + C - R^2 = 0$$

*we can find* $\Delta t$.

$$LifeTime[i,j] = \Delta t$$

$$\boldsymbol{LS[i,j]} = \frac{\boldsymbol{LifeTime[i,j]}}{\boldsymbol{\sigma}}$$

*Here,*

$LS[i,j] = 1$ *when* $LifeTime[i,j] \geq \sigma$

$\boldsymbol{LS_{i,j}}$ : *Link stability between two nodes i and j.*

Fig. 3 Reckoning Link Stability in EBGR

Once LS is calculated for each neighboring vehicle, EBGR selects the node corresponding to the highest LS (corresponding to the most stable neighboring link) as next hop for data forwarding. This approach should help as well in minimizing the risk of broken links and in reducing packet loss.

### 3.7 Potential Score Calculation (PS)

The potential score (PS) of all nodes present within the different levels of transmission range of source/packet forwarding node is calculated. The potential score (PS) is calculated to identify the closeness of next hop to destination, direction of motion of nodes and reliability of neighbor nodes. The appropriate edge node with largest potential score will be considered as having higher potential to reach the destination node and that particular node can be chosen as next hop to forward the packet to the destination node. Potential score is calculated by addition of DC, DMI and LS and that mathematical model represented in Fig.4.

$$\boldsymbol{PS_i} = \rho \times \mathbf{DC} + \omega \times \mathbf{DMI} + \lambda \times \mathbf{LS}$$

$$\boldsymbol{PS_i} = \rho \times \left(1 - \frac{\boldsymbol{D_i}}{\boldsymbol{D_c}}\right) + \omega \times \cos(\vec{v}_i, \vec{l}_{i,d}) + \lambda \times \boldsymbol{LS_{c,i}}$$

***Here,***

$\boldsymbol{PS_i}$ : *Potential score of node i*

$\rho, \omega, \lambda$ : *Potential factors*

*Let* $\rho + \omega + \lambda = 1$ ; $\lambda > \rho$ *and* $\lambda > \omega$

$\boldsymbol{D_i}$ : *Shortest distance from edge node i to destination D.*

$\boldsymbol{D_c}$ : *Shortest distance from packet forwarding node c to destination D.*

$\frac{\boldsymbol{D_i}}{\boldsymbol{D_c}}$ : *Closeness of nexthop.*

$\vec{v}_i$ : *Vector for velocity of edge node i.*

$\vec{l}_{i,d}$ : *Vector for the location of edge node i to the location of destination node D*

$\cos(\vec{v}_i, \vec{l}_{i,d})$ : *Cosine value of angle made by these vectors*

$\boldsymbol{LS_{c,i}}$ : *Link stability between packet forwarding node c to edge node i.*

Fig. 4 Potential Score Calculation in EBGR

### 3.8. Edge Node Selection (ENS)

In the Edge Node Selection, edge nodes are selected for packet forwarding event. An edge node is a node which has shortest distance to the destination D compared to all other nodes within the different levels of transmission range of source/packet forwarding node.

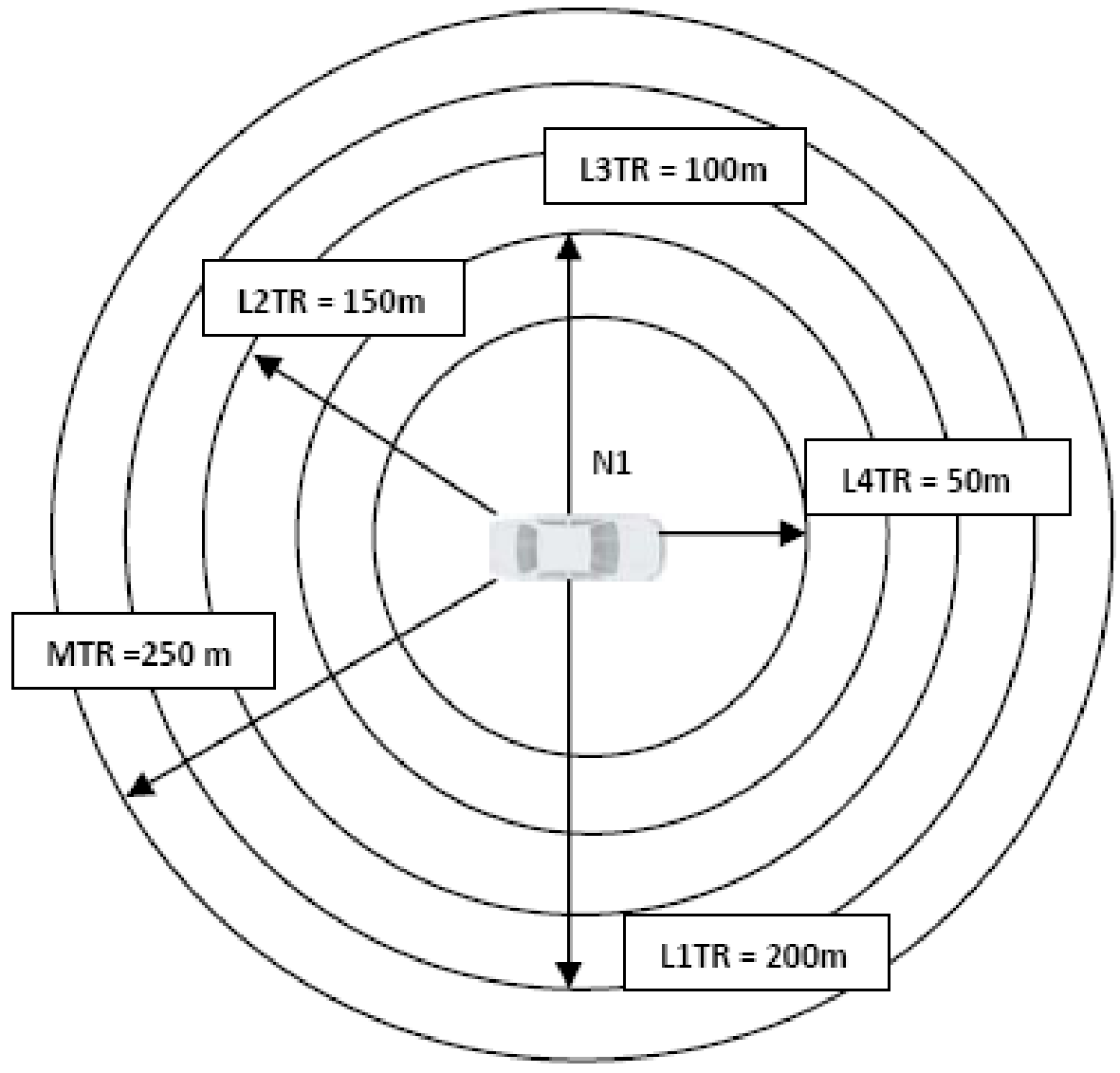

Fig 5: Different Levels of Transmission Range in EBGR

The different levels of transmission range are considered to avoid packet loss due to high speed mobility of vehicles. An edge node has the responsibility of saving received data packets in forwarding table and transfers it later when those nodes meet new neighbors. The overall objective of the algorithm is to forward the packet as soon as possible to increase packet delivery ratio, minimize the end to end delay and avoid packet loss. The MTR of a vehicle/node is 250m.The other levels of transmission range is considerably less than MTR. The different levels of transmission range is shown in Fig.5 which includes,

Maximum Transmission Range (i.e. MTR=250m)
Level1 transmission range (i.e.L1TR=200m)
Level2 transmission range (i.e.L2TR=150m)
Level3 transmission range (i.e.L3TR=100m)
Level4 transmission range (i.e.L4TR=50m).

$\boldsymbol{MTR}$: $Maximum\ Transmission\ Range = 250m$
$\boldsymbol{L1TR}$: $Level1\ Transmission\ Range = 200m$
$\boldsymbol{L2TR}$: $Level2\ Transmission\ Range = 150m$
$\boldsymbol{L3TR}$: $Level3\ Transmission\ Range = 100m$
$\boldsymbol{L4TR}$: $Level4\ Transmission\ Range = 50m$
$\boldsymbol{currentnode}$: $the\ current\ packet\ carrier$
$\boldsymbol{loc_c}$: $the\ location\ of\ current\ node$
$\boldsymbol{\vec{v_c}}$: $speed\ vector\ for\ current\ node$
$\boldsymbol{dest}$: $destination\ of\ the\ packet$
$\boldsymbol{loc_d}$: $the\ location\ for\ destination$
$\boldsymbol{nextHop}$: $the\ node\ selected\ as\ next\ hop$
$\boldsymbol{neigh_i}$: $the\ \boldsymbol{i}th\ neighbor$
$\boldsymbol{loc_i}$: $the\ location\ of\ the\ \boldsymbol{i}th\ neighbor$
$\boldsymbol{\vec{v}_i}$: $the\ speed\ vector\ of\ the\ \boldsymbol{i}th\ neighbor$

$loc_c \leftarrow getLocation(currentnode)$
$\vec{v_c} \leftarrow getSpeed(currentnode)$
$loc_d \leftarrow getLocation(dest)$
$D_c = distance(loc_c, loc_d)$
$\overrightarrow{l_{c,d}} = loc_d - loc_c$
$PS = \omega \times \cos(\vec{v_c}, \overrightarrow{l_{c,d}})$
$nextHop = currentnode$
$\textbf{for all}$ $neighbors\ of\ currentnode$ $\textbf{do}$
$loc_i \leftarrow getLocation(neigh_i)$
$\vec{v}_i \leftarrow getSpeed(neigh_i)$
$D_i = distance(loc_d, loc_i)$
$D_{ci} = distance(loc_c, loc_i)$
$\textbf{for all}$ $neighbors\ of\ currentnode\ with\ D_{ci}$ $\textbf{do}$
$\textbf{if}\ (D_{ci} < MTR\ \&\&\ D_{ci} > L1TR)$
$\vec{l}_{i,d} = loc_d - loc_i$
$PS_i = \rho \times \left(1 - \frac{D_i}{D_c}\right) + \omega \times \cos(\vec{v}_i, \vec{l}_{i,d}) + \lambda \times LS_{c,i}$
$\textbf{for}\ neigh_i\ with\ greater\ PS_i\ \textbf{do}$
$PS = PS_i$
$nextHop = neigh_i$
$\textbf{end for}$
$\textbf{else if}\ (D_{ci} < L1TR\ \&\&\ D_{ci} > L2TR)$
$\vec{l}_{i,d} = loc_d - loc_i$
$PS_i = \rho \times \left(1 - \frac{D_i}{D_c}\right) + \omega \times \cos(\vec{v}_i, \vec{l}_{i,d}) + \lambda \times LS_{c,i}$
$\textbf{for}\ neigh_i\ with\ greater\ PS_i\ \textbf{do}$
$PS = PS_i$
$nextHop = neigh_i$
$\textbf{end for}$
$\textbf{else if}\ (D_{ci} < L2TR\ \&\&\ D_{ci} > L3TR)$
$\vec{l}_{i,d} = loc_d - loc_i$
$PS_i = \rho \times \left(1 - \frac{D_i}{D_c}\right) + \omega \times \cos(\vec{v}_i, \vec{l}_{i,d}) + \lambda \times LS_{c,i}$
$\textbf{for}\ neigh_i\ with\ greater\ PS_i\ \textbf{do}$
$PS = PS_i$
$nextHop = neigh_i$
$\textbf{end for}$
$\textbf{else if}\ (D_{ci} < L3TR\ \&\&\ D_{ci} > L4TR)$
$\vec{l}_{i,d} = loc_d - loc_i$
$PS_i = \rho \times \left(1 - \frac{D_i}{D_c}\right) + \omega \times \cos(\vec{v}_i, \vec{l}_{i,d}) + \lambda \times LS_{c,i}$
$\textbf{for}\ neigh_i\ with\ greater\ PS_i\ \textbf{do}$
$PS = PS_i$
$nextHop = neigh_i$
$\textbf{end for}$
$\textbf{else if}\ (D_{ci} < L4TR)$
$\vec{l}_{i,d} = loc_d - loc_i$
$PS_i = \rho \times \left(1 - \frac{D_i}{D_c}\right) + \omega \times \cos(\vec{v}_i, \vec{l}_{i,d}) + \lambda \times LS_{c,i}$
$\textbf{for}\ neigh_i\ with\ greater\ PS_i\ \textbf{do}$
$PS = PS_i$
$nextHop = neigh_i$
$\textbf{end for}$
$\textbf{else}$
$carry\ the\ packet\ with\ currentnode$
$\textbf{end if}$
$\textbf{end for}$
$\textbf{end for}$

Fig. 6 Pseudo code of EBGR Algorithm

***Step1:*** Neighbor nodes having distance between 250m and 200m from the current node falls between MTR and L1TR. The potential score of all nodes present between the transmission range of MTR and L1TR are calculated. The node which is having higher potential score is considered as edge node of the MTR. So the packet from the current node is forwarded to that particular edge node. If no node present between MTR and L1TR, then L1TR and L2TR are considered.

***Step2:*** Neighbor nodes having distance between 200m and 150m from the current node falls between L1TR and L2TR. The potential score of all nodes present between the transmission range of L1TR and L2TR are calculated. The node which is having higher potential score is considered as edge node of the L1TR.So the packet from the current node is forwarded to that particular edge node. If no node present between L1TR and L2TR, then L2TR and L3TR are considered.

***Step3:*** Neighbor nodes having distance between 150m and 100m from the current node falls between L2TR and L3TR. The potential score of all nodes present between the transmission range of L2TR and L3TR are calculated. The node which is having higher potential score is considered as edge node of the L2TR.So the packet from the current node is forwarded to that particular edge node. If no node present between L2TR and L3TR, L3TR and L4TR are considered.

***Step4:*** Neighbor nodes having distance between 100m and 50m from the current node falls between L3TR and L4TR. The potential score of all nodes present between the transmission range of L3TR and L4TR are calculated. The node which is having higher potential score is considered as edge node of the L3TR.So the packet from the current node is forwarded to that particular edge node. If no node present between L3TR and L4TR, then L4TR are considered.

***Step5:*** Neighbor nodes having distance within 50m from the current node falls to L4TR. The potential score of all nodes present L4TR are calculated. The node which is having higher potential score is considered as edge node of the L4TR.So the packet from the current node is forwarded to that particular edge node. If no node present in any of the above mentioned range, then the current node store and carry the packet until it find some other node comes within its transmission range. The pseudo code of ENS algorithm is illustrated in Fig.6.

## 4. Simulation Results and Analysis

In this section, we evaluate the performance of routing protocols of vehicular networks in an open environment. So among the routing protocols we aforementioned, we choose GPSR, PDGR and EBGR for comparison.

### 4.1 Revival Mobility model (RMM)

We use Revival Mobility model (RMM) to simulate the movement pattern of moving vehicles on streets or roads defined by maps from the GPS equipped in the vehicles. In Revival Mobility model (RMM), the road comprises of two or more lanes. Vehicles or nodes are randomly distributed with linear node density. Each vehicle can move in different speed. This mobility model allows the movement of vehicles in two directions. i.e. north/south for the vertical roads and east/west for the horizontal roads. In cross roads, vehicles choose desired direction based on the shortest path. A security distance should be maintained between two subsequent vehicles in a lane. Overtaking mechanism is applicable and one vehicle can able to overtake the preceding vehicle. Packet transmission is possible and can be done by vehicles moving in both directions, which means front hopping and back hopping of data packet is possible as shown in the Fig.7.

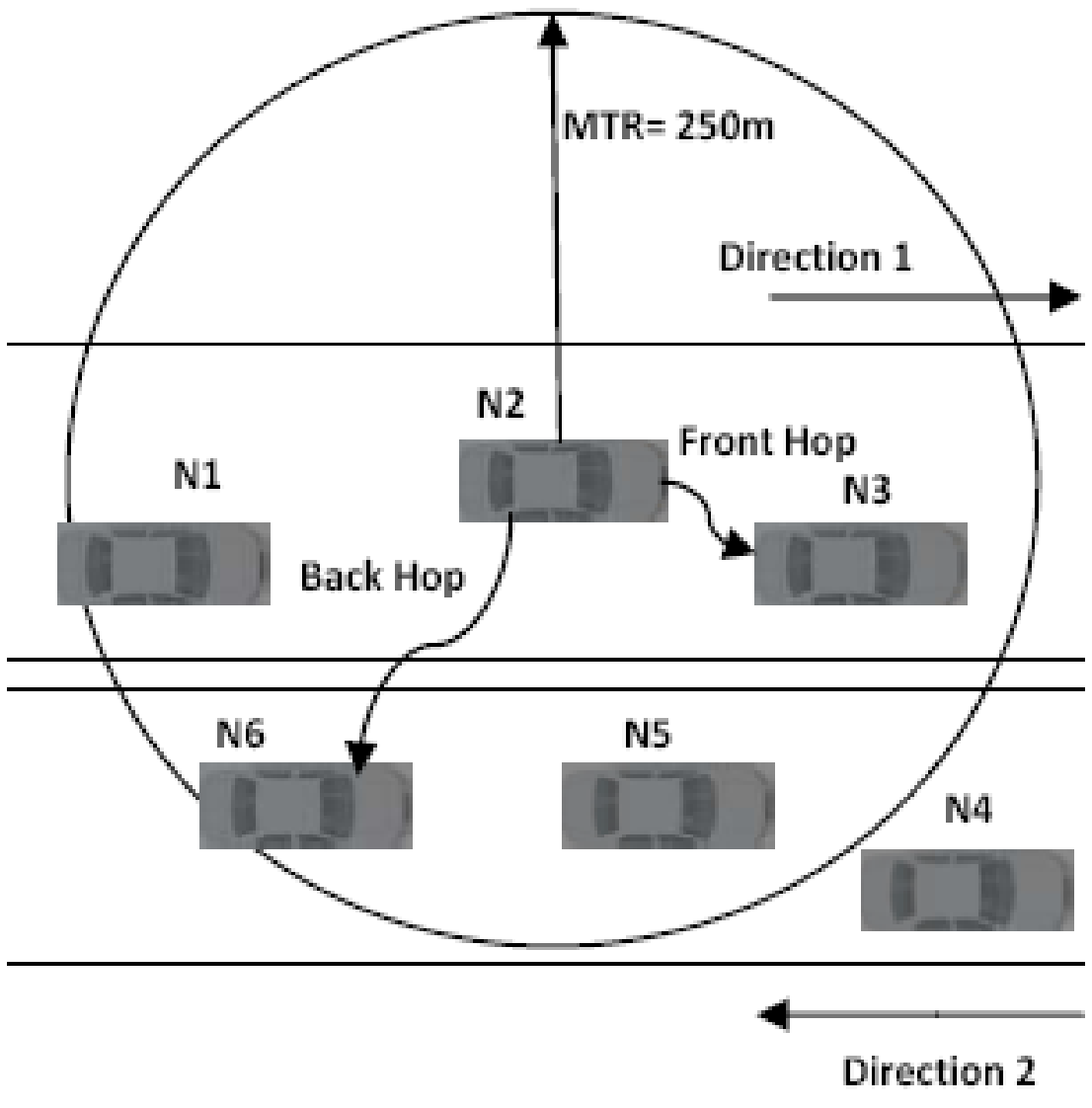

Fig.7 Revival Mobility Model

In this mobility, deterministic and instantaneous transmission mechanism in which a message is available for receiving within a certain radius $r=250m$ from the sender with certainty, but unavailable further away. Vehicles can unicast, multicast and broadcast packets to the neighbor vehicle which is present within its transmission range.

Table 2
Simulation Parameters

| Parameter | Value |
|---|---|
| Simulation Area | 1000m * 1000m |
| Number of Vehicles | 20 - 100 |
| Average speed of vehicles | 0 – 25 metre/second |
| Number of packet Senders | 40 |
| Transmission Range | 250m |
| Constant Bit Rate | 2 (Packets/Second) |
| Packet Size | 512 Bytes |
| Vehicle beacon interval | 0.5 (Seconds ) |
| MAC Protocol | 802.11 DCF |

The Simulations were carried out using Network Simulator (NS-2) ([20]). We are simulating the vehicular ad hoc routing protocols using this simulator by varying the number of nodes. The IEEE 802.11 Distributed Coordination Function (DCF) is used as the Medium Access Control Protocol. The packet size was fixed to 512 bytes. The Traffic sources are UDP. Initially the nodes were placed at certain specific locations, and then the nodes move with varying speeds towards new locations. The nodes move with speeds up to 25 meter/sec. For fairness, identical mobility and traffic scenarios were used across the different simulations. The simulation parameters are specified in Table 2

### 4.2. Performance Metrics to evaluate simulation

In order to evaluate the performance of vehicular ad hoc network routing protocols, the following metric is considered**.**

***Packet delivery ratio (PDR):*** The ratio of the packets that successfully reach destination.

$$PDR = \frac{Total\ number\ of\ packets\ \ delivered}{Total\ number\ of\ packets\ transferred} \times 100$$

### 4.3 Packet Delivery Ratio vs. Number of Nodes

In this part, we compare the packet delivery ratio with number of nodes and it is shown in Fig.8. Initially the packet delivery ratio is less for GPSR, PDGR and EBGR. When the number of node increases, then packet delivery for all routing algorithms increases. More nodes in network will provide more opportunities to find some suitable node for forwarding of packet. When no node available, GPSR switches to perimeter mode and it increases delay of packet transmission, which results in lower packet delivery ratio. PDGR have comparatively high packet delivery ratio compared with GPSR. PDGR considers its 2 hop neighbors for packet forwarding based on prediction. But prediction is not reliable at all situations. Due to high mobility, packet forwarded to edge of the transmission range will be lost. In EBGR, the next hop selection is done by considering the potential score. By using EBGR, the packet loss is minimized considerably and the packet delivery ratio is improved for about 11.6% in comparison with PDGR with the increase in number of vehicles.

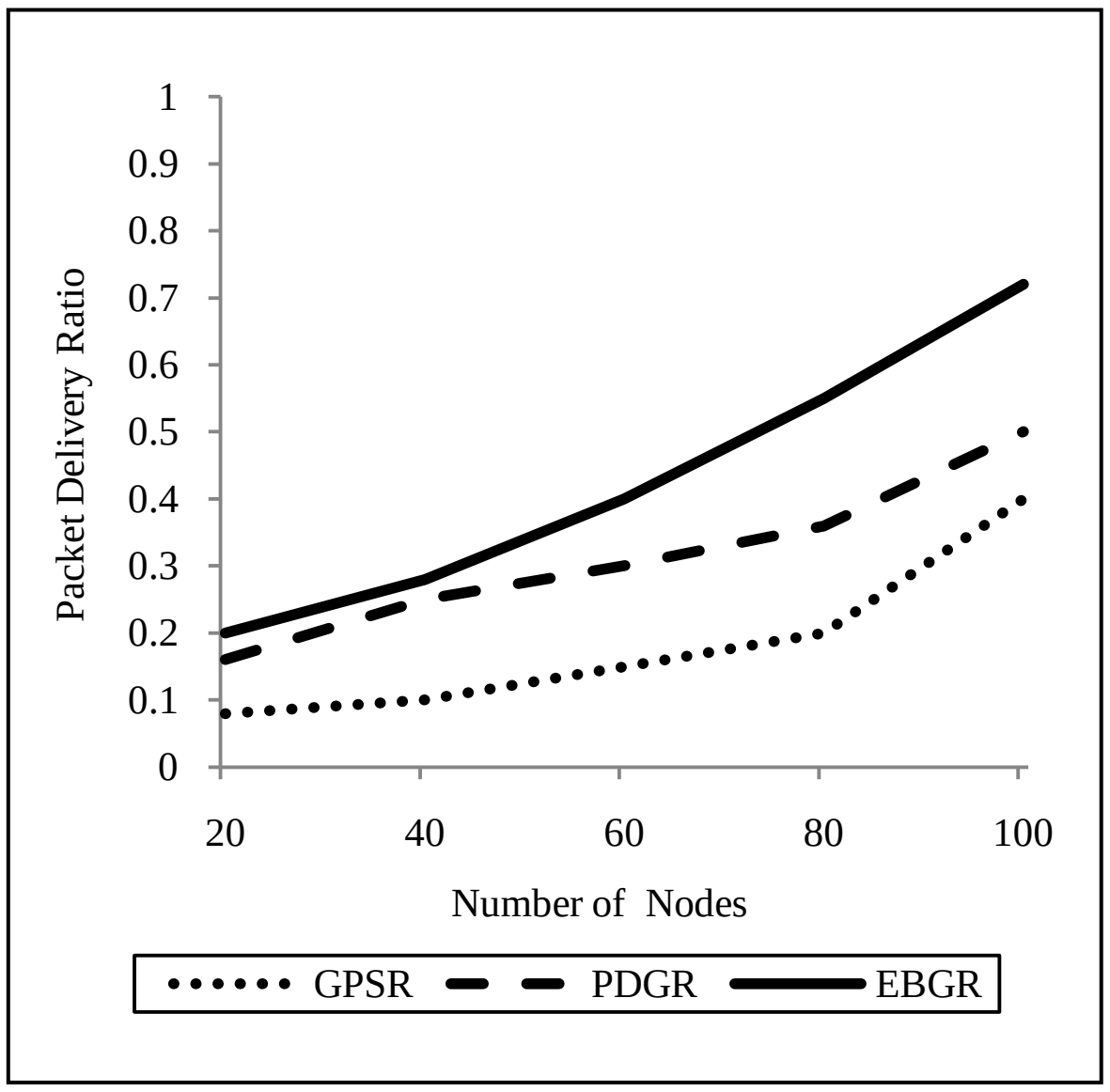

Fig. 8: Packet Delivery Ratio vs. Number of Nodes

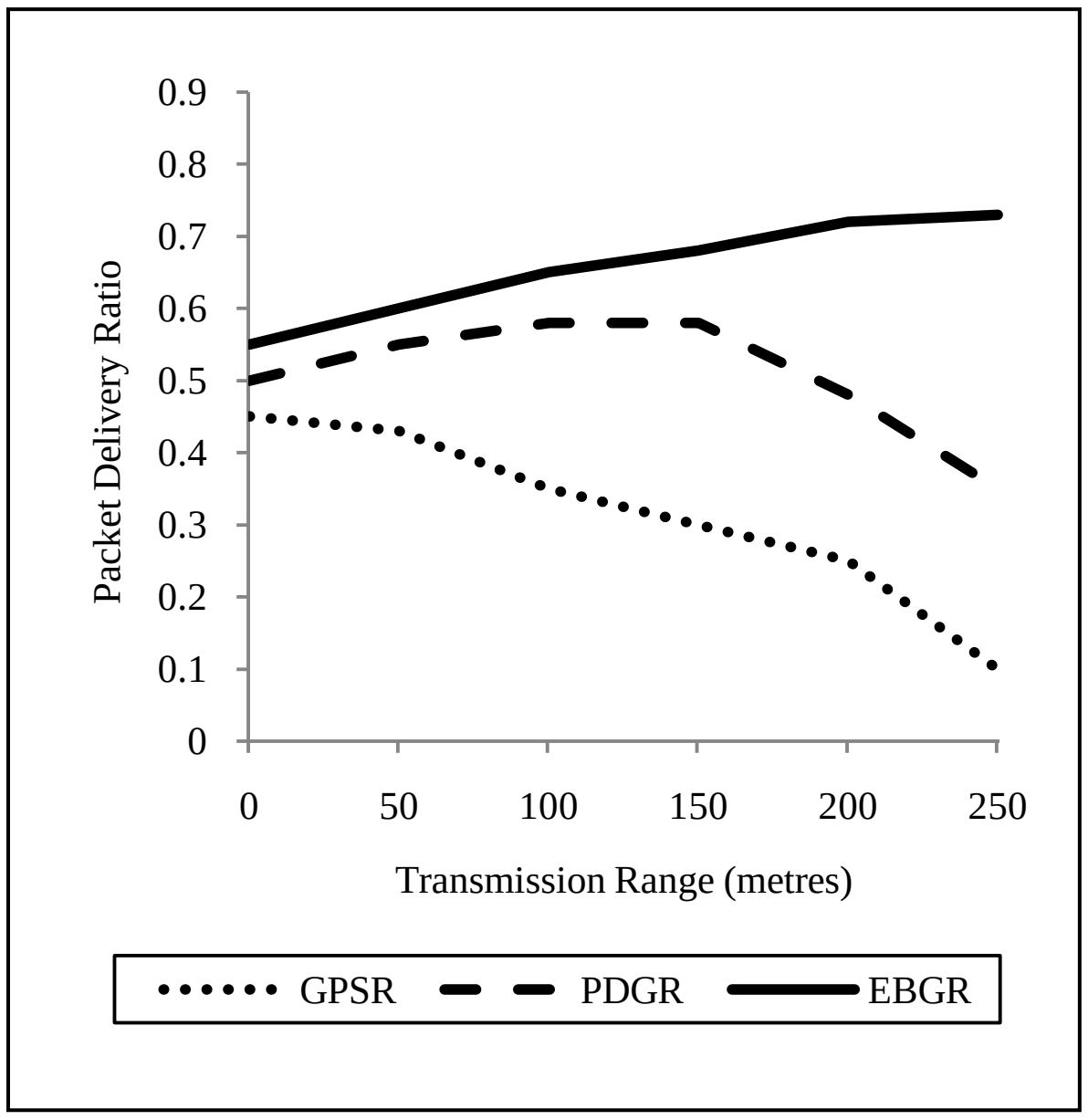

Fig. 9: Packet Delivery Ratio vs. Transmission range

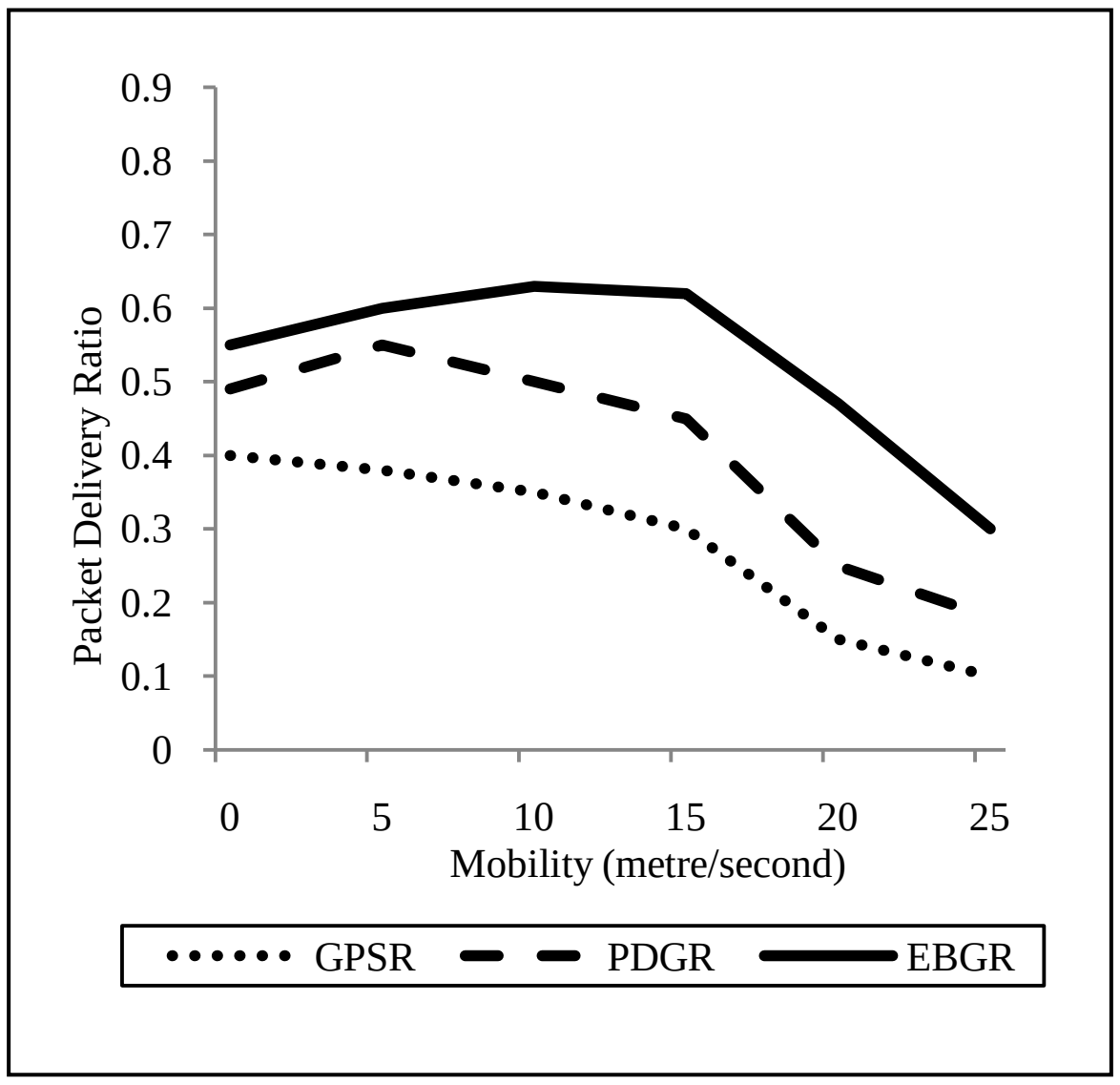

Fig. 10: Packet Delivery Ratio vs. Mobility.

### 4.4 Packet Delivery Ratio vs. Transmission Range

In this part, we compare the packet delivery ratio with different levels of transmission range and it is shown in Fig.9.The GPSR and PDGR always select the immediate one hop and two hop neighbors respectively to forward the packet. When many neighbor nodes are present, then numbers of hops are increased in GPSR and PDGR. This will decrease PDR. In EBGR, the vehicle always selects the neighbor node based on different levels of transmission range (i.e. L1TR, L2TR, and L3TR & L4TR) using distance information from GPS. By using EBGR, the numbers of hops are minimized considerably and the packet delivery ratio is improved for about 15.5% in comparison with PDGR with the different levels of transmission range.

### 4.5 Packet Delivery Ratio vs. Mobility

In this part, we compare the packet delivery ratio with varying speed of vehicles and it is shown in Fig.10. When the speed of vehicle increases, the packet delivery ratio of GPSR and PDGR decreases much faster than others. The high speed of vehicles leads to packet loss in edge of MTR. By using EBGR, the packet loss at the edge of MTR is minimized considerably and the packet delivery ratio is improved for about 12.5% in comparison with PDGR with the increase in speed of vehicles.

## 5. Conclusion

In this paper we have investigated routing aspects of VANETs. We have identified the properties of VANETs and previous studies on routing in MANETs and VANETs. We have commented on their contributions, and limitations. By using the uniqueness of VANETs, we have proposed Revival Mobility Model and a new position based greedy routing approach EBGR. Our simulation results have shows EBGR outperform GPSR and PDGR significantly in the terms of improving the packet delivery ratio. In the future, our approach requires modifications by taking into account the city environment characteristics and different mobility models with obstacles. Comparison of proposed EBGR approach with other existing approach shows that our routing algorithm is considerably better than other routing algorithms in improving the packet delivery ratio.

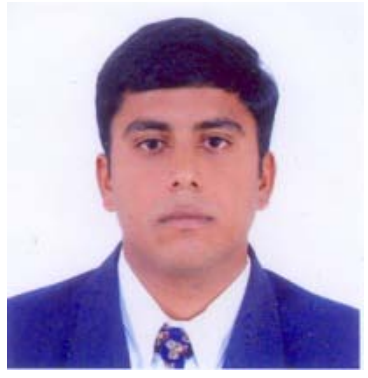

**K.Prasanth** received the B.E degree in Computer Science from K.S. Rangasamy College of Technology, Tiruchengode in 2005, M.Tech in Computer Science from SRM University, Chennai in 2007, MBA (System) in Periyar University, Salem in 2007. He worked as Project Engineer in Wipro Technologies, Bangalore from (2007-2008). He is currently working as lecturer in Department of Information Technology, K.S.Rangasamy College of Technology. His Current research interest includes Mobile Computing, Mobile Ad Hoc Networks and Vehicular Ad Hoc Networks.

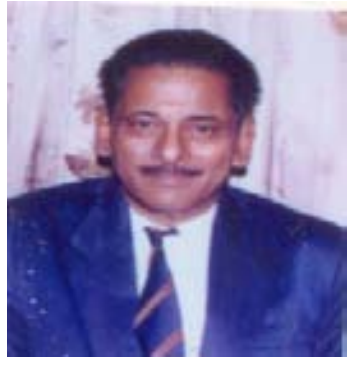

**Dr.K.Duraiswamy** received the B.E., M.Sc. and Ph.D. degrees, from the University of Madras and Anna University in 1965, 1968 and 1987 respectively. He worked as a Lecturer in the Department of Electrical Engineering in Government College of Engineering, Salem from 1968, as an Assistant professor in Government College of Technology, Coimbatore from 1983 and as the Principal at K.S.Rangasamy College of Technology from 1995. He is currently working as a Dean in the Department of Computer Science and Engineering at K.S.Rangasamy College of Technology (Autonomous Institution).His research interest includes Mobile Computing, Soft Computing, Computer Architecture and Data Mining. He is a senior member of ISTE, IEEE and CSI.

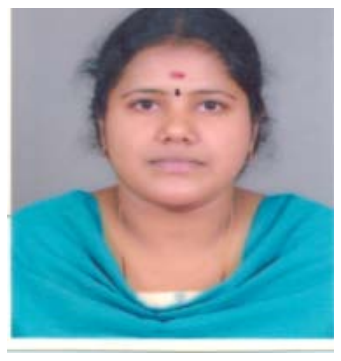

**K. Jayasudha** received the B.Sc degree in 2002, MCA degree in 2005, M.E degree in Computer Science in 2007, from K.S.Rangasamy College Of Technology, Thiruchengodu. She is currently working as Lecturer in Department Of Computer Applications, K.S.R.College Of Engineering, Thiruchengodu. Her current research interest includes Data Mining, Vehicular Networks.

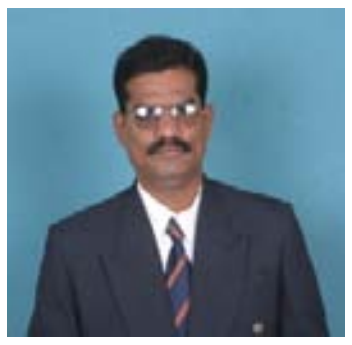

**Dr.C.Chandrasekar** received the B.Sc degree and M.C.A degree. He completed PhD in periyar university Salem at 2006. He worked as Head of the Department, Department Of Computer Applications at K.S.R.College of Engineering from 2007. He is currently working as Reader in the Department of Computer Science at Periyar University, Salem. His research interest includes Mobile computing, Networks, Image processing, Data mining. He is a senior member of ISTE, CSI.